\documentclass[12pt]{article}
\usepackage[english]{babel}

\usepackage{amsmath}
\usepackage{graphicx}
\usepackage[colorinlistoftodos]{todonotes}

\usepackage[font=scriptsize,labelfont=bf]{caption}
\usepackage{amsmath}

\usepackage{graphicx}
\usepackage{xcolor}
\usepackage{amssymb}
\usepackage{enumitem}
\usepackage{mathtools}

\usepackage{amsfonts}
\usepackage{amssymb}
\usepackage{amsbsy}
\usepackage{pifont}

\usepackage{hyperref}
\hypersetup{
    colorlinks,
    citecolor=black,
    filecolor=black,
    linkcolor=black,
    urlcolor=black
}
\usepackage{graphicx}
\graphicspath{ {images/} }

\usepackage[
backend=biber,
style=numeric,
sorting=none
]{biblatex}

\renewbibmacro{in:}{}

\usepackage[T1]{fontenc}

\usepackage{datetime}
\newdate{date}{18}{04}{2018}
\date{\displaydate{date}}

\begin{document} 

\begin{titlepage}

\newcommand{\HRule}{\rule{\linewidth}{0.5mm}} 

\center 
 

\textsc{\LARGE University College London}\\[1.5cm] 
\textsc{\Large Computer Science Department}\\[0.5cm] 
\textsc{\large A Literature Review\footnote{This report is submitted as part requirement for the module COMPGA11 Research in Information Security at University College London. It is substantially the result of our own work except where explicitly indicated in the text. The report may be freely copied and distributed provided the source is explicitly acknowledged.} on}\\[0.5cm] 


\HRule \\[0.4cm]
{ \huge \bfseries Privacy in Blockchain Systems}\\[0.4cm] 
\HRule \\[1.5cm]
 

\begin{minipage}{0.4\textwidth}
\begin{flushleft} \large
\emph{Author:}\\
Jad \textsc{Wahab} 
\end{flushleft}
\end{minipage}
~
\begin{minipage}{0.4\textwidth}
\begin{flushright} \large
\emph{Supervisors:} \\
Prof. George \textsc{Danezis} \\
\small{Alberto Sonnino \\
Mustafa Al Bassam}
\end{flushright}
\end{minipage}\\[2cm]



{\large \displaydate{date}}\\[2cm] 


\vfill 

\end{titlepage}

\addtocounter{page}{1} 


\setcounter{secnumdepth}{0} 

\begin{abstract}
In this literature review, we first briefly provide an introduction on the privacy aspect of blockchain systems and why it is a difficult quality to achieve, especially using traditional methods. Next, we go over a wide range of different strategies and techniques, along with their respective empirical implementations. Starting with approaches that attempted to provide privacy on Bitcoin/existing blockchain systems, then going into more advanced techniques, such as secure multi-party computations, ring signatures, and zero knowledge proofs, that construct a more advanced blockchain system from scratch with the objective of preserving privacy. Finally, we conclude that the current state of privacy on blockchains still needs work for it to be reliable. Nevertheless, the field of privacy in this domain is developing and advancing at a rapid rate.

\end{abstract}

\clearpage

\tableofcontents
\clearpage

\section{Introduction}

\subsection{Motivation and Goal}
Even though centralization has had its benefits in the past in terms of organization, efficiency, and speed, recent events have shown the weaknesses of centralization regarding corruption, abuse of power, and security (one point of failure). The original spirit of the Internet/cyberspace was one of fairness, freedom, and decentralization with the likes of cyber-libertarians such as John Barlow \cite{barlow_declaration_2016}. However, the Internet we now know today is far from that, dominated by large corporations and suffering from all types of manipulation, surveillance, and security breaches \cite{breaches} \cite{dominate_goog}.  \par

After the introduction of Bitcoin in 2009 \cite{nakamoto_bitcoin}, it opened the door for a lot of innovation and advancements in the world of applied cryptography and trust-less systems. Potential for more complex applications of blockchain technology was realized and people started adding to and evolving the technology in the space. A blockchain, ``a digital ledger in which transactions [or events] are recorded chronologically and publicly'' \cite{blockchain_dict}, allows for global interactions between parties in a trustless and secure manner that also leaves an audit trail that anyone can check/verify. \par

However, as promising as this technology may seem, it is not fully developed yet. There are yet many inherent shortcomings/difficulties; the main ones being privacy and scalability. By nature, fully decentralized, public, permissionless blockchains do not scale well and are pretty slow. In addition, blockchains and smart contract applications are visible to everyone and verifiable/runnable by everyone, making privacy a much harder characteristic to achieve in comparison to with traditional applications.

\subsection{Organization of the Review}
This literature review revolves around the academic as well and practical proposals and implementations that tackle the obstacle of privacy in blockchain systems. We first go over the earliest blockchain systems/platforms and how achieving privacy in such systems is a very difficult task due to the nature of how these systems operate. Following that, we touch upon the lessons learned from an experiment done which aimed at achieving privacy on such systems, specifically Ethereum. Next, we go through the main approaches that sought to accomplish this task, including direct approaches, secure multi-party computations, ring signatures, and zero knowledge proofs/techniques. Finally, we discuss the conclusions and outline future research directions and open problems of the field. Note that this review focuses mainly on public and permissionless blockchains.

\vfill
\section{Background}

\subsection{Blockchains \& Smart Contract Platforms}

\subsubsection{Bitcoin}
Even though Bitcoin is considered pseudo-anonymous and was used for anonymous payments on Silk Road \cite{silk_road}, it does in fact suffer from a considerable lack of privacy. Payment amounts are shown in the clear on the blockchain and transactions can be linked and traced all the way back to their origins. With the help of blockchain explorers such as Blockchain.info \cite{blockchain.info}, this process has become extremely easy, even for very amateur users.

\subsubsection{Ethereum}
In the early days of Bitcoin's existence, people had begun experimenting with Bitcoin's scripting language to establish more advanced functionality, such as colored coins \cite{colored_coins}. Buterin found the scripting language of Bitcoin too cumbersome to deal with and argued that Bitcoin needed a scripting language for application development. When he failed to get agreement from the Bitcoin developers/community, he established Ethereum, the first Turing-complete decentralized smart contract system \cite{eth_wp}. It is arguably the most prominent smart contract platform today, however, it is experiencing many difficulties that are preventing it from being used at a massive scale, such as privacy and scalability.

\subsubsection{Privacy-Preserving Smart Contract on Ethereum}
Real world privacy-preserving smart contracts/applications do not even have enough theoretical work, let alone any practical work. This can be mainly attributed to the fact that this whole area is very new and challenging. Unterweger et al. \cite{unterweger_lessons_2018} have actually implemented a privacy-preserving protocol from the energy domain as a smart contract, specifically on Ethereum. The system is currently functional but extremely infeasible and exhibits a large number of limitations. \par 
These limitations include: firstly, determining the gas limit for the nonrefundable fee spent to the miner is a difficult problem to solve on its own when considering both cost-effectiveness and throughput. The Ethereum blockchain inherently has trouble dealing with large data and transaction throughput delays, while privacy-preserving computations definitely come with their own overhead. Not to mention the fact that the Solidity programming language has shortcomings of its own including a very limited stack size and no native string handling functionality. But lastly, and most importantly, the resulting cost requirements are just plainly unacceptable by many orders of magnitude for many practical use cases. In the example by \cite{unterweger_lessons_2018}, deploying and executing the smart contract would cost around \$94, at the time of writing, where 1 ETH=\$460.75 \cite{ethereum_price}. \par
For any practical uses of privacy-preserving smart contract applications, there are still many limitations to overcome. State-of-the-art privacy enhancing technologies today cannot be used feasbily on blockchains that are not initially designed with privacy in mind.

\subsection{Traditional Methods for Privacy} 
\label{asdf}
Let us consider the ideal case, or ``holy grail,'' where users can do absolutely everything that they can do right now on a blockchain, but with the added feature of privacy, having the information completely obfuscated \cite{buterin_privacy_blog}. Unfortunately, it has been mathematically proven that perfect black-box obfuscation is impossible \cite{barak_impossibility}. So, in recent times, we have been settling for the next best thing, indistinguishablity obfuscation, which is still very powerful \cite{indis_obfus}. Indistinguishablity obfuscation is basically satisfying the condition that if we are given two different obfuscated programs that produce the same output, one cannot determine which of the two outputs original came from which source. This technology is used in software a lot nowadays, and especially in preventing malware from being detected. However, especially with the unscalable nature of blockchains, it does not mix well since ``the mechanism for doing this kind of obfuscation is horrendously inefficient; billion-factor overhead is the norm, and often even highly optimistic'' \cite{buterin_privacy_blog}.

\vfill
\section{Privacy on Blockchain Systems}
\subsection{Direct Approaches}
\subsubsection{CoinJoin}
CoinJoin is a Bitcoin anonymization technique proposed by Gregory Maxwell in August 2013 \cite{maxwell_coinjoin}. The basic idea of CoinJoin is to combine separate transactions together in the same transaction so that third parties cannot link the outputs with the inputs of the transaction. However, CoinJoin is subject to a number of weaknesses. First of all it is vulnerable to sybil or DoS attacks where malicious users send many sybil transactions to the mixer in order to eliminate them from the mix. Secondly, the whole method is very complex to implement well and is inefficient/slow. A good system would require proper timeout handling, proper security checks, and would need to be easy to use. A major drawback is that the system relies on other people and cannot be done alone. Plus, if one person times out, then the whole process fails. Finally, it does not provide guaranteed privacy and is not completely trustless (trusted mixer).

\subsubsection{Merge Avoidance}
Merge Avoidance was proposed by Bitcoin developer Mike Hearn in December 2013 \cite{hearn_merge_2013}. The technique is basically a  method to mitigate or reduce the likelihood of a third party being able to link addresses together on the blockchain. For example, if Alice needs 1 BTC from Bob, and he uses an unspent transaction output (UTXO) of 10 BTC to send her the 1 BTC and 9 BTC change back to himself, Alice will know that Bob has a minimum balance of 9 BTC. \par 
However, consider that Bob earns a 10 BTC salary from the coffee shop where he works. He could request his salary using the merge avoidance protocol. That way, instead of sending Bob a one-piece UTXO of 10 BTC, his employer would send him many different smaller UTXOs that he knows all of the private keys for. This is basically like asking for a payment in a change of small bills. Bob can even ask his employer to spread them out in order to prevent timing attacks. To a third party, they do not know that Bob is the recipient of all of these small transactions, until Bob later recombines the small outputs for a future payment. But Bob can also use CoinJoin after that to mix his coins, and so on the process continues. Evidently, the process is far from ideal and pretty messy with nothing close to approaching any real privacy guarantees.

\subsubsection{State Channels}
Using state channels is a technique where two parties send money/security deposit to a smart contract, then do whatever they want privately off-chain and then whenever they want to settle, they communicate with the blockchain. They can update states between themselves while introducing time locks and penalties in order to guarantee than no party maliciously commits an earlier state to the blockchain (similar to with the Lightning Network protocol \cite{poon_ln}).\par
In the context of running private code on smart contracts, they can include a \textit{hash} of code they want to run in a smart contract where they deposit money to, and then in the future when they want to settle, one of two cases can occur. In the first case, one party broadcasts the finalized state to the smart contract, the other party agrees, and the smart contract distributes the funds accordingly. In the other case however, one party broadcasts \textit{a} state to the smart contract and the other party disagrees/challenges by providing the input and the original code to the smart contract. Finally the smart contract executes the code (verifying its hash with the hash stored) with the input and distributes the funds accordingly \cite{buterin_privacy_blog}. In this case, privacy is only guaranteed when both parties cooperate. Plus, it has another implicit benefit in terms of scalability since a lot of work is done off-chain.

\subsubsection{Mobius}
Mobius is a system that replaces a centralized mixing service with an Ethereum smart contract that does the mixing autonomously. It combines the use of stealth keys and ring signatures to achieve anonymity, theft prevention, and low communication overhead (compared to similar previous systems). The autonomous mixing allows the system to resist availability attacks, where the centralized mixing service could just go offline after receiving the first payment(s)/before directing it, effectively stealing the money. It also achieves a much lower off-chain communication complexity than previous existing cryptocurrency tumblers, with senders and recipients needing to send only two initial messages in order to engage in an arbitrary number of transactions.

\subsection{Secure Multi-Party Computations}
Secure Multi-Party Computations (SMPCs) stem back to 1982 when Andrew Yao first proposed secure 2-party computation protocols in the same paper which included the famous \textit{millionaire's problem} \cite{yao_protocols_1982}. Even though this procedure is more efficient to the traditional method of indistinguishability obfuscation mentioned above, it still exhibits very large inefficiencies. \par
There are two main approaches in these private computations, homomorphic encryption and secret sharing. Homomorphic schemes are basically schemes in which computations can be done on ciphertexts in such a way that when the result is decrypted, it matches the result of the same operations done on the plaintexts. They have advantages in terms of simplicity and efficiency but exhibit issues in terms of expressiveness and mainly the question of ``who decrypts?'' since the result is the encryption of the output and not the output itself. On the other hand, secret sharing is where users split up their secret into ``shares'' sent across to different authorities who use specific protocols to perform functions on these shares without learning the secret in whole.

\subsubsection{Enigma}
Enigma is a ``decentralized computation platform with guaranteed privacy'' \cite{enigma} which is an implementation of SMPC based on secret sharing. At the time of writing (12/04/2018), the Enigma token (ENG) is worth \$127,303,811 in market cap with a ranking of $96^{th}$ worldwide \cite{enigma_price}. Basically, Enigma works by having the data split between nodes and having these nodes compute functions together but without leaking any information. Then Enigma off-loads any private or intensive computations off of the blockchain, and only has the the blockchain ``enforce access-control based on digital signatures and programmable permissions'' \cite{enigma}. Out of $n$ nodes that have shares of the secret, a threshold of $(t+1)$ nodes is needed to decrypt the secret and any subset of $t$ nodes cannot learn anything about the secret.\par 
For instance, suppose Alice wants to let Bob use her height in aggregate computations but without revealing it to him. She would invoke a private contract that includes her height (split in shares) which Bob can reference for computations, but \textit{cannot} access the height directly or know anything about the height unless he compromises at least $(t+1)$ nodes. \par
The consensus of the system is reached through an incentives based threefold mechanism relying on: security deposits, computation fees, and storage fees. Not much is discussed in the paper about this mechanism but it does raise some questions regarding the technique to preventing the lack of guaranteed fairness in the protocol as well as overall complexity and vague process of realizing/penalizing malicious nodes. Not to mention how the computation fees are calculated beforehand since the platform is turing-complete so how is a node to know what the minimum threshold for a request is before accepting it?\par
In addition, there are a lot of complications with multiplication computations in the system. First of all, even though work has been done to reduce latency from quadratic growth to linear growth (at the cost of increased computation complexity which is parallelized), it still includes a lot of latency and complexity. It also reduces security from $t$ nodes to $\frac{t}{2}$ nodes when the requirement of $t$ node security trust is in itself a huge requirement/constraint. Last but not least, there is no way to tell if any nodes maliciously colluded so it is impossible to incentivise honesty in such a system. As a result, SMPC systems like this are arguably more suited to private blockchains which are inherently faster and exhibit external incentives to guard against such security constraints.

\subsection{Ring Signatures}
Ring signatures are, in essence, signatures which prove that a signer has a private key belonging to a set of different private keys but without revealing \textit{which one} of them it is. \par
Linkable ring signatures are a specific form of ring signatures where a ring signature can only be used once by each of the members of the set, or in other words, to prevent the ``double spending'' problem. The double spending problem is inherent to digital currencies and applications since they are not physical objects and is basically the problem of preventing a token from being spent more than once to more than one person. Outside of the context of digital currencies, this can be used in e-voting schemes for example, where each voter from the set can only vote once.
\subsubsection{CryptoNote}
The CryptoNote protocol seeks to utilize ring signatures in order to provide \textit{unlinkability} and \textit{untraceability} properties to systems like Bitcoin. Unlinkability is where, unlike Bitcoin, even if one's receiving address is revealed, transactions sent to this address cannot be linked to it. This is done by the use of one-time stealth addresses for each transaction. Untraceability is where outputs of a transaction cannot be traced back to it's inputs. This is done using ring signatures on the outputs of the transaction. \par
\noindent Considering the parameters below, the protocol works as follows: \par
q: a prime number: $q = 2^{255} - 19$ \par
d: an element of $\mathbb{F}_q$: d = $\frac{-121665}{121666}$ \par
E: an elliptic curve with equation: $-x^2 + y^2 = 1 + dx^2y^2$ \par
G: a base point; $G = (x, \frac{-4}{5})$ \par
l: a prime order of the base point: $l = 2^{252} + 27742317777372353535851937790883648493$ \par
$\mathcal{H}_s$: a cryptographic hash function $\{0,1\}* \xrightarrow{} \mathbb{F}_q$ \par
$\mathcal{H}_p$: a deterministic hash function $E(\mathbb{F}_q) \xrightarrow{} E(\mathbb{F}_q)$ \\

\noindent First of all, instead of having only one Elliptic Curve (EC) key-pair like Bitcoin, every user has two EC key-pairs: 
$$(a, A) \text{ and }(b, B) \mid a,b \in [1, l-1], a\neq b, A=aG, B=bG$$ 
One of the key-pairs is basically used for spending and the other key-pair for viewing. A transaction happens by first having the sender perform a non-interactive Diffie-Hellman exchange \cite{diffiehellman} in order to generate the one time stealth shared secret for the stealth address. \par  
To do this, the sender generates a random $r \in [1, l-1]$ and computes the one time stealth address using the public keys of the receiver $(A,B)$: 
$$P = \mathcal{H}_s(rA)G+B$$ 
The sender then sends the transactions to $P$ but also includes $R=rG$ in the transaction so that the receiver can use it to determine the Diffie-Hellman exchange key-pair. \par
Since only the sender and receiver know $r$, only they can know who the transaction was sent to. Then, the receiver checks all transactions with his private keys $(a,b)$ and the $R$ value to compute:
$$P = \mathcal{H}_s(aR)G+B$$ 
If P'=P, then the receiver will know that that transaction is meant for him since the Diffie-Hellman exchange will be completed. Finally, when the receiver of this transaction wants to spend it, they can recover the corresponding one-time private key:
$$p = \mathcal{H}_s(aR)+b \mid P=pG$$ 

The two main implementations of CryptoNote as of writing this paper are ByteCoin and Monero, and are often used in conjunction with an I2P (Invisible Internet Project) routing service called \textit{Kovri} or The Onion Route (TOR) to provide location (IP) anonymity as well. Moreover, the anonymity guarantees  provided by CryptoNote implementations and the like are not as completely strong as they seem since some information can still be recovered \cite{DBLP:journals/corr/MillerMLN17}. For example, even though the outputs of the transaction are fragmented and untraceable, if at a later stage, they are linked to the same transaction, it can be inferred that they all belonged to the same user. Note that many adjacent transactions can be done to further obfuscate this process in order to hamper the detection/linking process mentioned even more. Furthermore, Monero uses key images to prevent double spending and keep track of which outputs are spent and which are not. This undoubtedly makes scaling attempts and objectives much more difficult to achieve.

\subsection{Zero Knowledge Proofs/Techniques}
Zero Knowledge Proofs (ZKPs) are basically proofs/methods where one party is able to prove knowledge of something to another party \textit{without} revealing anything about it to that person (for more clarification see interesting practical example of colorblind friend in \cite{zkp_ex}). This cryptographic technique paved the way for many advancements in the world of applied cryptography. With regards to blockchain systems such as Bitcoin, the idea is that there is actually no need for the accounts and balances to be visible to everyone, they just have to be \textit{verifiable} by everyone in order to prevent the ``double spending'' problem. Cryptographers took this idea and worked on systems such as the ones mentioned below.

\subsubsection{Zcash/PinocchioCoin/Zcoin}
Zcoin is the first implementation of the Zercoin protocol implemented by \cite{zerocoin}. The original Zerocoin protocol relies heavily on the ``Strong RSA assumption'' and the double-discrete logarithm proof. Zcoin tackled privacy issues in terms of linkability but still reveals payment destinations and amounts, and is limited in functionality. It also includes old and established techniques with known performance restrictions. Because of this, Pinocchio Coin \cite{danezis_pinocchio_2013} was proposed that uses instead elliptic curves and bi-linear pairings. It also utilizes smaller proofs and quicker verification as a result of using modern techniques based on quadratic arithmetic programs. \par
Zcash was built on the Zerocash protocol \cite{zerocash} which, even though was proposed by mostly the same authors of Zerocoin, operated differently. It relied on utilizing the recent advancements made in Zero Knowledge Succinct Non-interactive ARguments of Knowledge (ZK-SNARKs). ZK-SNARKs are basically zero knowledge proofs of knowledge that are extended to prove validity of \textit{any} function (as opposed to just knowledge of something). In my opinion, it is struggling to gain adoption nowadays since it does not inherently implement privacy, is too slow, and very complicated. Most transactions are almost identical to Bitcoin and called ``transparent'' transactions. Private transactions are called ``shielded'' transactions and make up only around 13\% of the total transactions on the network \cite{davitt_zcash_2018}. Even though verification of the shielded transactions takes milliseconds, they take around a minute to create and require 8GB of RAM, eliminating the ability to be done on modern smartphones.

\subsubsection{Hawk}
Proposed in mid 2016, Hawk \cite{kosba_hawk} is a framework for building privacy-preserving smart contracts. It can be seen as an extension of the work done by Zerocash  and such but which also addresses programmability (which has been forgone in those systems). The system works as follows. A non-specialist programmer is able to write a Hawk program in the typical form without having the implement any cryptography and the Hawk compiler complies the program into three parts: the blockchain's program which will be executed by all consensus node, the program to be executed by the users, and the program to be executed by a special facilitating party called the manager. These parts jointly define a cryptographic protocol between three entities: the users, the manager, and the blockchain. This ``minimally trusted'' manager does not have the ability to influence/affect the correct execution of the contract, but must be trusted not to disclose users' private data. In a nutshell, this system can be seen as a system where the blockchain is trusted for authenticity and only one specific party (manager) is trusted for privacy. Nonetheless, the system does not have a running implementation as of the time of writing nor is it optimized for scalability.

\subsubsection{Chainspace}
Chainspace is a recently proposed, sharded smart contract platform with privacy built in by design \cite{al-bassam_chainspace:_2017}. As opposed to the account/state based model of Ethereuem, Chainspace uses an object model. Everything in Chainspace is expressed in the form of an object (that can possess several attributes) which is either \textit{active} or \textit{inactive}. Once an object is set as inactive, it can never go back to being active. This can be seen as analogous to the UTXO Bitcoin model of unspent and spent transactions but in a more general than just currency transactions. In order to update the state of an object, a new object must be created with the new state and the object having the old state will become inactive. \par
Unlike Ethereum, Chainspace implements privacy by design as follows. There are two parts of Chainspace smart contracts: the user side (executor) and the node side (checker). On the user side, the user takes the previous state of the smart contract along with the private data they have and executes the smart contract locally to create the output state. Then, the user sends that output state to the node side. On the node side, instead of re-running/executing the code like in Ethereum, all the node has to do is verify the execution in zero knowledge (without learning anything about the private data of the user). This can be illustrated with a very simplified example of a smart contract with the functionality of creating digital signatures. On the user side, the user executes the smart contract to create a signature using his \textit{private} key. Next the user sends the digital signature, which is essentially the output state of the contract to the node. On the node side, all the node has to do is verify that the digital signature is valid. This simple example can be extended to more advanced cases with more advanced smart contract functionality. Note that the protocol for the execution of the smart contract determines the generation of the zero knowledge proof to be sent to the node who will use it to assert the correctness of the encryption and thus the execution. \par
To enable scalability on Chainspace, the nodes are organized into shards that manage the state of objects, keep track of their validity, and record transactions aborted or committed. The nodes ensure that only valid transactions, consisting of encrypted or committed data, along with the zero knowledge proofs that assert their correctness, end up on their shard of the blockchain. The nodes also communicate with the other shards to decide whether to accept or reject a transaction (intra-shard consensus). To do this, Chainspace proposes a protocol called \textit{Sharded Byzantine Atomic Commit} or $\mathcal{S}$-BAC, which basically combines existing Byzantine agreement and atomic commit primitive protocols in a novel way. \textit{Byzantine agreement} securely keeps consensus on a shard of $3f+1$ nodes in total, containing up to $f$ malicious nodes. \textit{Atomic commit} runs across all shards that contain objects which the transaction relies on. The transaction is rejected unless all of the shards accept to commit the transaction. The process is shown in \textbf{figure 1} below.\par

\begin{figure}[h]
\centering
    \includegraphics[scale=0.4]{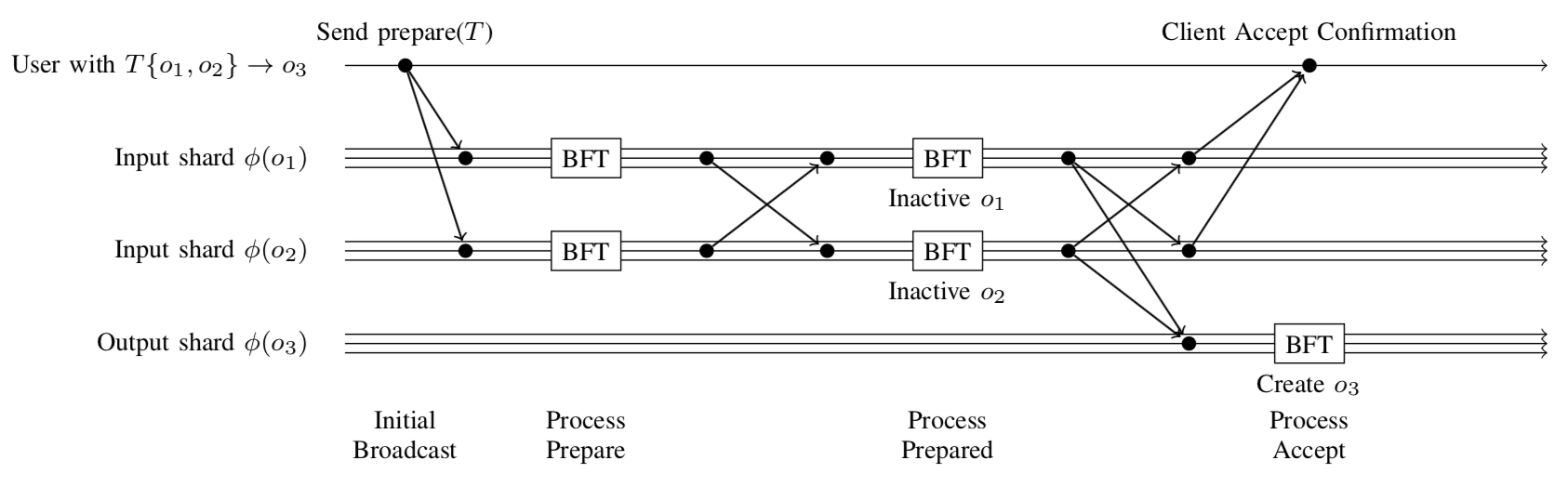}
    \caption{
    $\mathcal{S}$-BAC for a transaction T with two inputs $(o1,o2)$ and one output object $(o3)$. The user sends the transaction to all nodes in shards managing o1 and o2. The BFT-Initiator takes the lead in sequencing T , and emits `prepared(accept, T)’ or `prepared(abort, T)’ to all nodes within the shard. Next the BFT-Initiator of each shard assesses whether overall ‘All proposed(accept, T)’ or `Some proposed(abort, T)’ holds across shards, sequences the accept(T,*), and sends the decision to the user. All cross-shard arrows represent a multicast of all nodes in one shard to all nodes in another.}
\end{figure}

The security properties obtained by this system include: transparency, encapsulation, integrity, and non-repudiation. Transparency is achieved since anyone can verify the authenticity of the history of transactions. Encapsulation is achieved since smart contracts do not interfere with objects of other smart contracts, unless it is defined by that smart contract. Integrity is achieved since if the malicious nodes do not exceed $f$, only valid transactions will be committed to the blockchain. Non-repudiation is achieved since misbehavior is detecatable on the network. Not to mention, unlike in Hawk where there is a  trusted third party (TTP), the \textit{manager}, which must be trusted not to disclose private data, in Chainspace this is not the case as the system does not rely on any TTPs.

\vfill
\section{Conclusions and Future Directions}
There are still a lot of challenges that need to be faced before we can be satisfied with the level of privacy achieved on blockchains and smart contract applications. Nevertheless, we are in the right direction since development and advancements are occurring very rapidly as this review demonstrates. Another issue hindering the growth of blockchain technology is evidently scalability. If solutions to scalability on the blockchain are produced, there is no doubt that it would help advance the privacy enhancing technologies being developed on the blockchain as well.\par
In addition, another thing we must keep in mind is that, similar to what is happening today with encrypted data on the internet, even though \textit{contents} of the data are private, it does not necessarily guarantee the privacy of the \textit{meta-data}, especially if there is enough to be statistically analyzed.

\clearpage
\section{References}
\AtNextBibliography{\small}
\printbibliography[heading=none]

\end{document}